\documentclass[conference]{IEEEtran}
\IEEEoverridecommandlockouts
\usepackage{cite}
\usepackage{amsmath,amssymb,amsfonts}
\usepackage{algorithmic}
\usepackage{graphicx}
\usepackage{textcomp}
\usepackage{xcolor}
\usepackage{subfigure}
\usepackage{fancyhdr}
\usepackage[normalem]{ulem}
\usepackage{url}
\usepackage{tikz}
\usepackage{soul, color, xcolor}
\usepackage{amsthm}
\usepackage{amsmath,amsfonts}
\usepackage{algorithmic}
\usepackage{graphicx}
\usepackage{textcomp}
\usepackage[table,xcdraw]{}

\usepackage{booktabs} % For formal tables
\usepackage{stfloats}
\usepackage[normalem]{ulem}
\usepackage{subfigure}
\usepackage{multirow}
\usepackage{paralist}
\usepackage{balance}
\usepackage{bm}
\usepackage[ruled,linesnumbered]{algorithm2e}
\usepackage{diagbox}
\usepackage{threeparttable}
\usepackage{makecell}
\usepackage{colortbl}
\usepackage{bm}
\usepackage{color}
\usepackage{tabularray}

\usepackage{lipsum}
\usepackage[compact]{titlesec}

\newtheorem{definition}{Definition}
\DeclareMathDelimiter{(}{\mathopen} {operators}{"28}{largesymbols}{"00}
\DeclareMathDelimiter{)}{\mathclose}{operators}{"29}{largesymbols}{"01}

\definecolor{grey}{RGB}{0.5,0.5,0.5}
\usepackage{tcolorbox}

\def\BibTeX{{\rm B\kern-.05em{\sc i\kern-.025em b}\kern-.08em
    T\kern-.1667em\lower.7ex\hbox{E}\kern-.125emX}}
\begin{document}

% \title{FIXME: End-to-End Benchmarking of LLM-Aided Design Verification}
\title{FIXME: Towards End-to-End Benchmarking of LLM-Aided Design Verification
% should not be used}How LLMs Are Redefining End-to-End Repair in Hardware Design  Fine-Grained Evaluation of LLM-Aided Hardware Design  FIXME: A Comprehensive Framework for Evaluating LLM Capabilities in Hardware Design Refinement FIXME: On the Fine-Grained Evaluation of LLM-Aided Hardware Design Refinement
% \thanks{Identify applicable funding agency here. If none, delete this.}On the Evaluation of End-to-End Repair Ability for LLM-Aided Hardware Design
}

\author{
\IEEEauthorblockN{Gwok-Waa Wan$^{1,2}$, Shengchu Su$^{1}$, Ruihu Wang$^{2}$, Qixiang Chen$^{2}$, Sam-Zaak Wong$^{1}$, Mengnv Xing$^{2}$,\\ Hefei Feng$^{1}$, Yubo Wang$^{2}$, Yinan Zhu$^{2}$, Jingyi Zhang$^{1}$, Jianmin Ye$^{1}$, Xinlai Wan$^{2}$, Tao Ni$^{2}$, \\Qiang Xu$^{2,3}$, Nan Guan$^{4}$, Zhe Jiang$^{1,2}$, Xi Wang$^{1,2}$, and Yang Jun$^{1,2}$}
\IEEEauthorblockA{$^1$ School of Integrated Circuits, Southeast University, Nanjing, Jiangsu, China}
\IEEEauthorblockA{$^2$ National Center of Technology Innovation for Electronic Design Automation, Nanjing, Jiangsu, China}
\IEEEauthorblockA{$^3$ The Chinese University of Hong Kong, Hong Kong SAR, China}
\IEEEauthorblockA{$^4$ City University of Hong Kong, Hong Kong SAR, China}
\IEEEauthorblockA{$^*$ Corresponding author: xi.wang@seu.edu.cn}
}

\maketitle

\begin{abstract}
Despite the transformative potential of Large Language Models (LLMs) in hardware design, a comprehensive evaluation of their capabilities in design verification remains underexplored. Current efforts predominantly focus on RTL generation and basic debugging, overlooking the critical domain of functional verification, which is the primary bottleneck in modern design methodologies due to the rapid escalation of hardware complexity. We present FIXME, the first end-to-end, multi-model, and open-source evaluation framework for assessing LLM performance in hardware functional verification (FV) to address this crucial gap. FIXME introduces a structured three-level difficulty hierarchy spanning six verification sub-domains and 180 diverse tasks, enabling in-depth analysis across the design lifecycle. Leveraging a collaborative AI-human approach, we construct a high-quality dataset using 100\% silicon-proven designs, ensuring comprehensive coverage of real-world challenges. Furthermore, we enhance the functional coverage by 45.57\% through expert-guided optimization. By rigorously evaluating state-of-the-art LLMs such as GPT-4, Claude3, and LlaMA3, we identify key areas for improvement and outline promising research directions to unlock the full potential of LLM-driven automation in hardware design verification. The benchmark is available at https://github.com/ChatDesignVerification/FIXME.
\end{abstract}


\section{INTRODUCTION}
\label{sec:INTRO}
The integration of Large Language Models (LLMs) into hardware design workflows, termed LLM-Aided Design (LAD), is emerging as a transformative paradigm in Very Large Scale Integration Circuit (VLSI) design methodologies~\cite{fu2023gpt4aigchip}. This integration promises to revolutionize hardware engineering through enhanced agile design processes and intelligent iteration cycles. The potential impact of LAD extends beyond mere automation, indicating a fundamental reshaping of how we approach hardware design challenges in the coming decades.

Recent advances have demonstrated the feasibility of LAD across Register Transfer Level (RTL) generation to chip-level planning. Notable achievements include the successful fabrication of chips designed with significant LLM assistance~\cite{wang2024chatcpu}, enabling automated coding, debugging, assertion insertion, random stimulus generation, and design optimization~\cite{LLMAssertions ,xu2024meic,liu2023verilogeval,LLM4SecHW}. This paradigm shift is particularly crucial given the exponentially increasing complexity of modern hardware systems, where traditional design methodologies are struggling to keep pace with development demands and the need for seamless collaboration among engineering teams.

\begin{figure}[!t]
        \centering
        \includegraphics[width=0.5\textwidth]{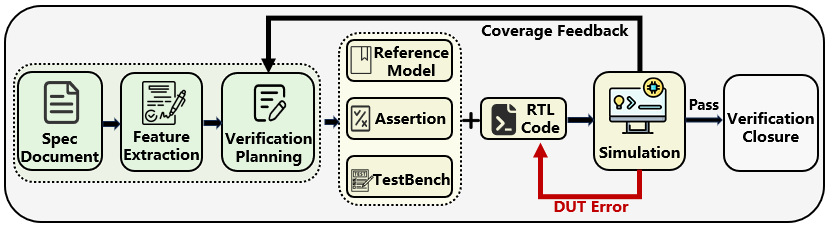}
        \caption{Traditional Design Functional Verification~\cite{DeliveringFunctionalVerification}}
        \label{fig:sec1:flow}
\end{figure}

\begin{figure*}[!t]
    \centering
    \begin{minipage}{0.37\textwidth} % Adjust width to fit layout
        \centering
        \includegraphics[width=\textwidth,height=3cm]{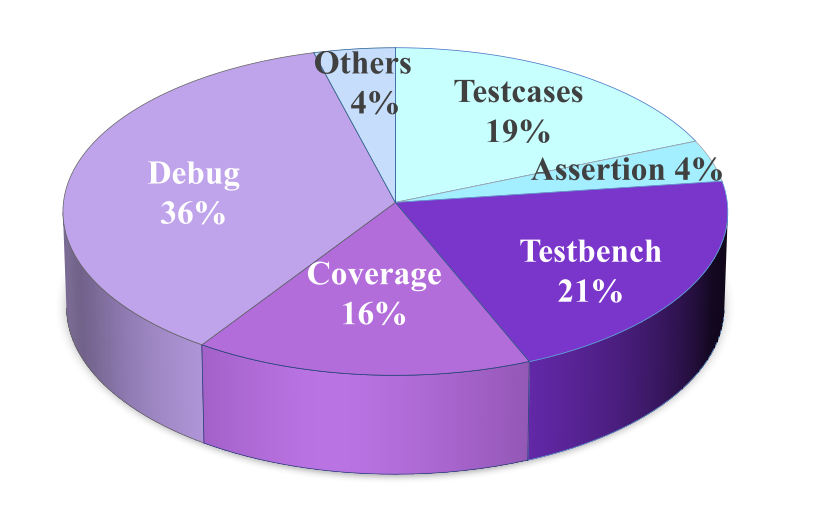}
        \caption{Time Spent by Sub Domains of Verification~\cite{DeliveringFunctionalVerification}}
        \label{fig:sec1:time}
    \end{minipage}
    \hspace{0.05\textwidth} % Adjust space between subfigures if needed
    \begin{minipage}{0.53\textwidth}
        \centering
        \includegraphics[width=\textwidth]{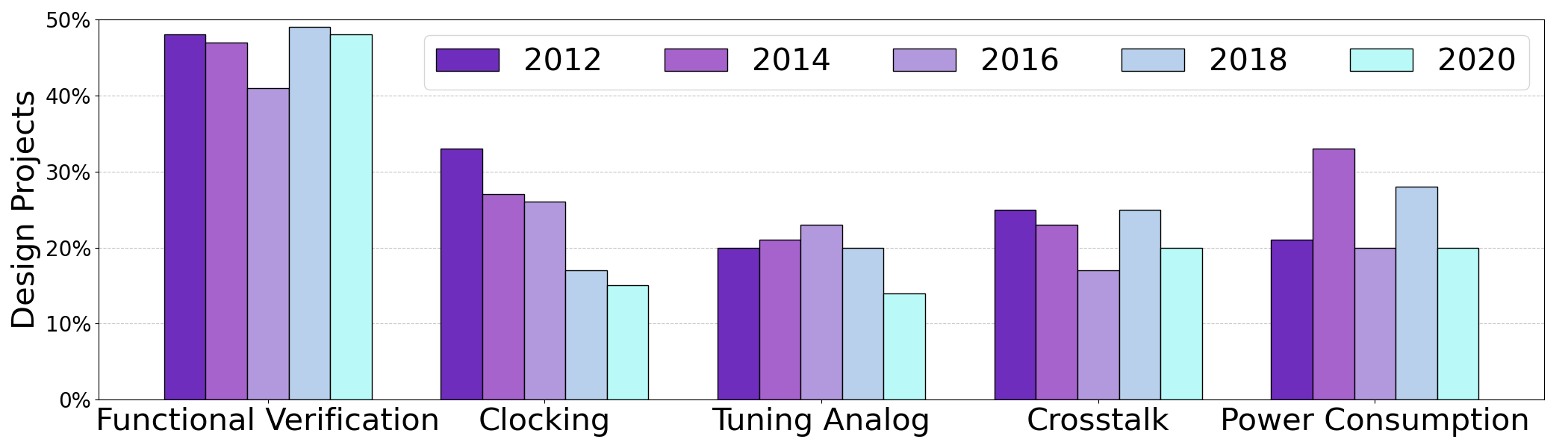}
        \caption{Top Reasons for Silicon Failed~\cite{10.1145/3638046}}
        \label{fig:sec1:fail}
    \end{minipage}
\end{figure*}

However, a significant aspect is lagging: most existing works focus on code generation rather than comprehensive test acceleration. The more critical bottleneck in agile hardware development lies in functional verification (FV)~\cite{Falsafi2023AgileVerification}. Figure~\ref{fig:sec1:flow} shows the general process of FV, which now accounts for up to two-thirds of the entire design cycle~\cite{8356004}, and approximately half of the design issues that result in silicon re-spins can be attributed to it,  as shown in Figure~\ref{fig:sec1:fail}. 

Given this context, the significance of advancing LLM research in FV becomes evident. A fine-grained evaluation of FV capabilities is both vital and urgent for guiding future LAD research directions. Unfortunately, existing benchmarks for LAD primarily emphasize code generation~\cite{fu2023gpt4aigchip, Chip-Chat},  while testbench tasks or formal verification are rarely considered. This scope of assessment fails to capture the full breadth and complexity of real-world FV challenges, which encompass crucial tasks such as specification comprehension, feature extraction, testbench generation, reference model design, SystemVerilog assertion (SVA) creation, and in-depth debugging. The proportion of time spent on these tasks in FV cannot be ignored, as Figure~\ref{fig:sec1:time} shows.

To address these critical limitations, we present \textbf{FIXME}, the first open-source evaluation framework for LLM-aided design functional verification. The structured approach of FIXME to assessment, spanning multiple difficulty levels and verification sub-domains, will provide insights into the role of LLMs in enhancing and automating the FV workflow. By establishing a standardized, industry-aligned evaluation platform, we aim to catalyze advancements in this rapidly evolving field and empower hardware designers to navigate the increasing complexity of modern design with greater confidence and efficiency through LLM cooperation. This research makes the following key contributions.

\begin{itemize}
\item \textbf{First Functional Verification Benchmark:} To our best knowledge, \textbf{FIXME} is the first end-to-end, multi-modal, and quantitative evaluation framework for LAD FV. It covers various sub-domains, including specification comprehension, model generation, testbench design, assertion, and debugging. This comprehensive approach ensures reliable and seamless testing of the responses of LLMs tailored for real-world-level FV tasks.
\item \textbf{Task Diversity:} FIXME unravels the entire hardware verification flow by introducing a structured three-level difficulty hierarchy, encompassing six sub-domains and 180 tasks, which encompass a diverse range of tasks from multiple-choice questions to verification code generation. This granular approach enables fine-grained evaluation.
\item \textbf{Robust Evaluation Dataset:} FIXME constructs a high-quality dataset using an AI-human collaborative approach. We design a multi-agent system, VerifyAgent, for accelerating benchmark design on identifying, classifying and filtering tasks. Coupled with expert-refined reference answers and quantitative metrics, this method ensures coverage of real-world verification scenarios and benchmark completion at an efficient speed. This AI-human flow also can be transformed for agile benchmark construction of other data-lack domains.
\item \textbf{Expert-Enhanced Function Coverage: } For the 180 tasks involved in FIXME, based on the experience of industrial-level experts, we have optimized the specification of 58K tokens, audited 15K lines of design code, supplemented 25K lines of testbench, added 1,235 crucial assertions, and established a unified interface for the reference model. Consequently, the functional coverage has increased by an average of 45.57\%. All of the above are available with open-source access.
\item \textbf{Systematic Analysis and LAD Guidance:} Through a comprehensive evaluation of state-of-the-art (SOTA) LLMs using FIXME, we identify areas for improvement and outline promising future research directions. By discovering the critical need for LLM-aided design verification, we hope that FIXME can be one of the guidelines for future work enhancement.
\end{itemize}

This paper is organized as follows. Section 2 reviews related work on LLM-aided hardware design and existing benchmarks. Section 3 details the design of our proposed benchmark, including the selection of hardware designs, the definition of tasks, and evaluation metrics. Section 4 describes the experimental setup, including the LLM models and evaluation environment. Section 5 presents the experimental results, analyzing the performance of different LLMs on various tasks. Section 6 concludes the paper.

\section{Background}
\label{sec:background}
%Section 2. Background
\subsection{Functional Verification}
The functional verification (FV) is a crucial phase in the hardware design process, primarily aimed at ensuring that the design meets its specification requirements and operates as intended~\cite{Wang2009Verification}. It aims to identify design flaws early in the development cycle, preventing costly silicon re-spins. Generally, the verification of RTL correctness is categorized into two primary methodologies: simulation verification and formal verification~\cite{Girish2021Verification, pixley2003functional, Lam2008Verification}. Simulation verification is a dynamic approach that validates the design by simulating its behavior under various input conditions. While it provides a flexible and intuitive means of verification, it often relies heavily on the expertise of engineers and can be inefficient, particularly for complex designs. In contrast, formal verification is a static method that employs mathematical techniques to prove the correctness of a design~\cite{Kern1999FormalVerification,Bening2001FormalVerification}. This method does not depend on specific input values; instead, it abstracts and reasons about the design to verify whether it satisfies certain properties. Although formal verification is precise, it typically incurs higher learning and implementation costs. In practice, a combination of both methods is commonly employed to enhance the efficiency and reliability of the verification process. Regardless of the approach, the increasing complexity of designs significantly escalates the time and resources required for verification~\cite{xu2022towards, Coussy2010hls}. In contemporary chip development, functional verification accounts for as much as 70\% of the front-end development timeline, making agile verification methodologies a necessity~\cite{pixley2003functional, Rashinkar2001verification, verigen}.

\subsection{LLM-Aided Design and Benchmarks}
In recent years, LLMs have achieved remarkable advancements in natural language processing, and their capabilities are progressively being applied to hardware design. Various initiatives, such as VeriGen~\cite{verigen}, BetterV~\cite{BetterV}, and RTLCoder~\cite{RTLCoder} focus on Hardware Description Language (HDL) code generation, while tools like RTLFixer~\cite{RTLFixer} and MEIC~\cite{xu2024meic} illustrate the application of LLMs for RTL debugging. Other projects, including ChipChat~\cite{Chip-Chat}, ChipGPT~\cite{chang2023chipgpt}, GPT4AIGChip~\cite{fu2023gpt4aigchip}, and ChatCPU~\cite{wang2024chatcpu} explore the acceleration of overall chip design through LLM integration. Additionally, LLM4DV~\cite{zhang2023llm4dv}, VerilogReader~\cite{VerilogReader}, AssertLLM~\cite{AssertLLM}, and AutoBench~\cite{AutoBench} investigate the role of LLMs in design verification. 

Benchmarking has emerged as a primary means to evaluate model capabilities, with ongoing developments in this area. Currently, most researches are concentrated on HDL code generation, exemplified by initiatives like RTLLM~\cite{RTLLM}, VerilogEval~\cite{liu2023verilogeval}, and VeriGen~\cite{verigen}. In contrast, benchmarks specifically targeting verification are still in their infancy. For instance, FVEval~\cite{kang2024fveval} focuses on evaluating formal verification, but there remains a blank in benchmarking that addresses real-world functional verification requirements.

\subsection{Gaps in Current Research}
\label{sub-domain}
As previously noted, existing works predominantly focus on design generation or performance optimization. While these works provide valuable insights into the capabilities of HDL generation, they often offer only superficial assessments of verification tasks. This misalignment creates an imbalance in the evaluation process, leading to situations where LLMs may excel on certain benchmarks for code generation yet falter when confronted with the complexities inherent in real-world hardware design verification scenarios.

This discrepancy underscores a critical gap in the current research landscape, highlighting an urgent need for a comprehensive evaluation framework that specifically addresses the practical application of LLMs in the context of functional verification. Such a framework is essential to ensure 1) Content that covers both simulation and formal verification; 2) A structural design that includes task classification and difficulty management to reflect model performance at a granular level; 3) A process that incorporates the major sub-domains in functional verification, including design specification comprehension, model generation, functional point design, testbench generation, assertion generation, and debugging for the Design Under Test (DUT).

\section{DESIGN \& METHODOLOGY}
\label{sec:design}
.
% 第三章: 
% 题目规划（定义＋抽象）
% 定义＋数学空间假设，作为序言，说明规划好了fixme的基本涵盖类型；
% 每种类型需要不同的题型方式，兼顾了评测有效性、效率和客观性；

% 框架设计（模组化可拓展）
% 构建了不同子域相互解耦、题目与评测流程解耦的模组化框架，供后续的侧重不同子域和题目更新的benchmark研究拓展

% 数据构建（ai＋human）

% 指标formalism （每个子域考察什么指标。以及如何量化）$\mathcal{S}$   $\mathcal{M}$  $\mathcal{R}$   $\mathcal{O}$  $\mathcal{I}$    $\mathcal{\tau}$  $\mathcal{T}$
% $\mathcal{A}$  $\mathcal{D}$
%

To address the research gaps discussed in Section 2, we propose the FIXME, a comprehensive benchmark covering the entire FV flow below processes.

\subsection{Preliminaries}
We formally define the key components of the FV process.

\begin{definition}[Specification ($\mathcal{S}$)]
\label{def:spec}
A gold-standard document $\mathcal{D} = {r_1, r_2, ..., r_n}$ comprising a set of requirements $r_i$ that completely characterize the intended functionality and the behavioral constraints of the hardware design. Serves as the axiomatic basis for FV~\cite{gajski1995specification}.
\end{definition}

\begin{definition}[Reference Model ($\mathcal{M}$)]
\label{def:refmodel}
A high-level abstraction typically implemented in SystemC/C++ that provides a golden behavioral implementation $\mathcal{M : I \rightarrow O}$ mapping input space $\mathcal{I}$ to output space $\mathcal{O}$ according to $\mathcal{S}$. Provides expected outputs for RTL equivalence checking through simulation~\cite{clarke2003behavioral}.
\end{definition}

\begin{definition}[RTL Implementation ($\mathcal{R}$)]
\label{def:rtl}
A design abstraction that models a synchronous digital circuit in terms of the flow of digital signals between hardware registers and the logical operations performed on those signals.
Typically described in hardware description languages (HDL) like Verilog or VHDL~\cite{vahid2010digital}.
\end{definition}

\begin{definition}[Testcase ($\mathcal{\tau}$)]
\label{def:testcase}
A complete stimulus-response tuple $\tau = (i_\tau, o^{expected}_\tau)$, where $i_\tau \in \mathcal{I}$ denotes input vectors and $o^{expected}_\tau  \in \mathcal{O}$ the denotes expected outputs derived from $\mathcal{S}(i)$. Serves to verify specific functional features~\cite{bergeron2007writing}. 
\end{definition}

\begin{definition}[Testbench ($\mathcal{T}$)]
\label{def:testbench}
A SystemVerilog/UVM verification environment comprising: 1). stimulus generators and clock/reset controllers; 2).
scoreboards for output validation; 3). functional coverage monitors and 4). interface adapters for DUT integration~\cite{spear2008systemverilog}. 
\end{definition}

\begin{definition}[Assertion ($\mathcal{A}$)]
\label{def:assertion}
A temporal logic property expressed in SystemVerilog Assertions (SVA) that specifies invariant behavioral constraints derived from $\mathcal{S}$~\cite{foster2004assertion}. For an \textbf{assertion} $a \in \mathcal{A}$, must hold $\forall t \in \mathbb{T}$ during simulation/formal FV:
\begin{equation}
    \alpha: \mathcal{D}(s_t) \rightarrow \{0,1\}
\end{equation}
where $s_t$ is the design state at time $t$.
\end{definition}

\begin{definition}[Debug ($\mathcal{D}$)]
\label{def:debug}
 The process of identifying, isolating, and correcting design errors. It involves analyzing simulation waveforms, log files, and code to pinpoint the root cause of failures~\cite{sarangi2006cadre}.
\end{definition}

\subsection{Problem Formulation}

We abstracted the FV process into a discretely observable set of states. We assumed the following scenario. 

\textbf{FV Formulation.} Given a design specification $\mathcal{S}$ described in natural language, the objective is to obtain a reference model $\mathcal{M}$, a group of testcases $\mathcal{\tau}$, a testbench $\mathcal{T}$, and a set of assertions $\mathcal{A}$ based on the understanding of $\mathcal{S}$. The testbench $\mathcal{T}$ and assertions $\mathcal{A}$ can be considered collections of verification primitives, where $\mathcal{T}$ contains a set of testcases $\mathcal{\tau}$ and $\mathcal{A}$ comprises basic assertion statements $a$. The verification process is then carried out by applying simulation tools, denoted as the verification function $\mathcal{F}$. The testbench $\mathcal{T}$ and assertions $\mathcal{A}$ are utilized to validate the RTL code $\mathcal{R}$, which can be expressed as ${\Delta = F(R, \tau, a, M)}$, where $\forall \tau$ in $T$, a in $\mathcal{A}$. The verification outcome $\Delta$ is either a "pass" or "fail" result. In the event of a failed verification, the process of debugging, refining, and correcting the RTL code $\mathcal{R}$ based on the specification $\mathcal{S}$ must be undertaken until an equivalent version $\mathcal{R'}$ is achieved.

% 框架设计（模组化可拓展）
% 构建了不同子域相互解耦、题目与评测流程解耦的模组化框架，供后续的侧重不同子域和题目更新的benchmark研究拓展

% 数据构建（ai＋human）

% 指标formalism （每个子域考察什么指标。以及如何量化）$\mathcal{S}$   $\mathcal{M}$  $\mathcal{R}$   $\mathcal{O}$  $\mathcal{I}$    $\mathcal{\tau}$  $\mathcal{T}$
% $\mathcal{A}$  $\mathcal{D}$

\subsection{Design Philosophy} 

%以上讨论和分析展示出FV是一项长链的、涉及到多个阶段的工作，每个阶段的领域知识、底层能力息息相关，任何一个阶段的问题都有可能导致验证收敛放缓。为此，高效的FV总是端到端对齐且全流程高效的。显然，综合评估LLM在FV中的能力，将有助于发现模型验证的边界，定位提升空间。这引出了FIXME的设计哲学：立足FV的全流程，端到端的实现评估，兼顾评测有效性、题目客观性和任务的多样性，尽量在丰富的设计类型和验证需求中需求可量化的平衡。

%为此，我们根据FV的一般过程，提取出了关键步骤，按照一一对应的原则，设计了具体的任务分类，一共六类，如表~\ref{表：示例}所示，包含了设计理解(SC)、参考模型生成（MG）、基本测试用例设计(TD)、完整的testbench生成（TG）、assertion生成（AG）和debug(Debug)。为了适合定量分析，我们对每种类型的任务进行了适当的调整: 1）.SC为客观题目，根据提供的设计规范进行选择；2）. MG为代码生成题目，根据测试结果判断语法和功能通过性；3）. TD为简答题，根据语义相似度进行判别；4）. TG为代码生成题目，根据仿真结果获取语法、功能正确性，以及line、toggle coverage等主要指标；5）. AG为代码生成题，通过仿真结果判断assertion的语法和功能正确性；以及6）. Debug为联合调试题目，以仿真结果是否通过作为判别。通过这样的调整，FIXME初步实现了可量化。

%FIXME将六类题目划分成为6个不同的子域。每个子域包含30道测试题目，整个benchmark共计180道测试题目。考虑到benchmark的可拓展性和验证工作的多样性，我们将FIXME设计成解耦式架构，即不同子域相互解耦、题目与评测流程解耦，供后续的侧重不同子域和题目更新的benchmark研究进行定制化拓展。

\textbf{End-to-end Strategy.} FV represents a complex, multi-stage process where domain-specific knowledge and underlying capabilities are intricately interconnected. Challenges at any single stage can potentially decelerate verification convergence. Consequently, efficient FV necessitates an end-to-end aligned and holistically optimized workflow. The comprehensive evaluation of LLMs in FV serves a critical purpose: identifying model verification boundaries and locating potential improvement opportunities. This motivation underpins the design philosophy of FIXME: establishing a comprehensive, end-to-end assessment framework that balances evaluation effectiveness, task objectivity, and task diversity across rich design types and verification requirements.

\textbf{Diversified Taxonomy.} Inspired by the general FV workflow, we extracted key procedural steps and designed corresponding task classifications following a one-to-one mapping principle. The framework encompasses six distinct verification sub-domains, as delineated in Table 1: specification comprehension (SC), reference model generation (MG), basic testcase design(TD), testbench generation (TG), assertion generation (AG), and debugging (Debug). To facilitate quantitative analysis, each task type was appropriately adjusted as follows: 1) SC tasks are objective questions requiring selection based on provided design specifications; 2) MG tasks involve code generation, with syntax and functional correctness evaluated according to test outcomes; 3) TD tasks are short-answer questions assessed through semantic similarity measures; 4) TG tasks entail code generation, with evaluation criteria including simulation results for syntax, functionality, as well as line and toggle coverage metrics; 5) AG tasks consist of code generation, with assertion syntax and functional correctness verified via simulation results; and 6) Debug tasks are judged by whether simulation results pass. These adjustments enable quantifiable assessment for FV tasks.

\textbf{Decoupled Architecture.} FIXME categorizes the six task types into six distinct sub-domains, each containing 30 test items, resulting in 180 tasks for the entire benchmark. To ensure extensibility and accommodate diverse verification needs, the benchmark adopts a modular architecture. In this design, sub-domains are mutually decoupled, and tasks are separated from the evaluation workflow, thereby enabling customized expansion and updates focused on specific sub-domains or task types in future benchmark studies.
%%%%%%%%%%%%%%%%%%%%%%%%%%%%%%%%%%%%%%%%%%%%%%%%%%%%%%%%%%%%%%%%%%%%%%%%%%%%%%%%%%%%%%%%%%%%%%%%%%%%%%%%%%

\begin{table}
\label{Table:FV2FIXME}
\caption{Main Functional Verification Stages and Related Tasks in FIXME.} 
     \vspace{-2pt}
\centering
\resizebox{\linewidth}{!}{
\begin{tabular}{lllll} 
\hline
Categories                                                    & Stage in FV                                                   & Tasks in FIXME                                                        & Type                                                               & Count  \\ 
\hline
\begin{tabular}[c]{@{}l@{}}Design\\Comprehension\end{tabular} & \begin{tabular}[c]{@{}l@{}}Feature \\Extraction\end{tabular}  & \begin{tabular}[c]{@{}l@{}}Specification \\Comprehension\end{tabular} & \begin{tabular}[c]{@{}l@{}}Multiple\\Choice \\Question\end{tabular} & 30      \\ 
\hline
\begin{tabular}[c]{@{}l@{}}Design \\Modeling\end{tabular}      & \begin{tabular}[c]{@{}l@{}}Reference \\Model\end{tabular}     & \begin{tabular}[c]{@{}l@{}}Model \\Generation\end{tabular}            & Coding                                                             & 30      \\ 
\hline
\begin{tabular}[c]{@{}l@{}}Basic \\Test\end{tabular}    & \begin{tabular}[c]{@{}l@{}}Testcase\\Design\end{tabular}      & \begin{tabular}[c]{@{}l@{}}Testcase \\Design\end{tabular}     & \begin{tabular}[c]{@{}l@{}}Short \\Answer\end{tabular}             & 30      \\ 
\hline
Testbench                                                     & \begin{tabular}[c]{@{}l@{}}Simulation \\with TB\end{tabular}  & \begin{tabular}[c]{@{}l@{}}Testbench \\Generation\end{tabular}        & Coding                                                             & 30      \\ 
\hline
Assertion                                                     & \begin{tabular}[c]{@{}l@{}}Formal \\Verification\end{tabular} & \begin{tabular}[c]{@{}l@{}}SVA \\Generation\end{tabular}              & Coding                                                             & 30      \\ 
\hline
Debug                                                         & Debug                                                         & \begin{tabular}[c]{@{}l@{}}DUT Debug\end{tabular}                & Coding                                                          & 30      \\
\hline
\end{tabular}}
\end{table}
%%%%%%%%%%%%%%%%%%%%%%%%%%%%%%%%%%%%%%%%%%%%%%%%%%%%%%%%%%%%%%%%%%%%%%%%%%%%%%%%%%%%%%%%%%%%%%%%%%%%%%%%%%%%

\begin{table*}
\centering
\caption{Examples and Metrics of Tasks in FIXME.} %\textcolor{red}{check the  descriptions and examples as I don't have domain knowledge}}
     \vspace{-2pt}
     \label{Table:example}
\centering
\resizebox{\linewidth}{!}{
\begin{tblr}{
  width = \linewidth,
  colspec = {Q[125]Q[265]Q[420]Q[128]},
  hlines,
}
Tasks in FIXME              & Target                                                                     & Example                                                                                                                                                            & Metrics                                     \\
Specification  Comprehension \textbf{(SC)} & Evaluating the ability of the model to interpret design specifications.         & Which of the following \textcolor[rgb]{0,0.392,0}{statements correctly} describe the functionality and structure of the "\textit{hpdmc\_mgmt}" module?             & Pass Rate                                          \\
{Model
\\Generation \textbf{(MG)}}        & Evaluating the generation ability of reference models.        & According the spec and RTL, generate a \textcolor[rgb]{0,0.392,0}{reference model} for co-simulation for module "\textit{i2c\_slave}".                              & Syntax; Functional Pass Rate                          \\
{Testcase Design \textbf{(TD)}} & Assessing the capacity of the model to identify key DUT features for FV.                            & Design the \textcolor[rgb]{0,0.392,0}{testcase} for signal "\textit{pc\_next}".                                                                                    & Semantic Completeness~                                    \\
{Testbench
\\Generation  \textbf{(TG)}}    & Assessing the creation of simulation-based verification scripts.           & According the spec and RTL, generate a \textcolor[rgb]{0,0.392,0}{testbench} for module "\textit{writeqspi}".                                                       & {Syntax; Functional; \\Line\textbackslash{}Toggle Cov.} \\
{SVA
\\Generation \textbf{(AG)}}          & Evaluating the design of formal verification primitives.                   & Design the \textcolor[rgb]{0,0.392,0}{assertions} to verify the write address signal (h\_waddr) computed correctly in the module "\textit{equivalence\_resolver}". & Syntax; Functional Pass Rate              \\
{DUT/TB
\\Debug \textbf{(Debug)}}            & Assessing the refinement and fixing of issues based on simulation results. & Please \textcolor[rgb]{0,0.392,0}{ fix the errors }in the "\textit{fpu\_add}" module with the help of Spec and error log: "\textit{line 27, syntax error: ......}"  &Functional Pass Rate                  
\end{tblr}}
\end{table*}

%%%%%%%%%%%%%%%%%%%%%%%%%%%%%%%%%%%%%%%%%%%%%%%%%%%%%%%%%%%%%%%%%%%%%%%%%%%%%%%%%%%%%%%%%%%%%%%%%%%%%%%%%%%%

\subsection{Constructing FIXME: Challenges and Solutions}
The development of a robust and balanced task dataset is critical for the construction of effective benchmarks. After confirming the assessment content, we encountered several challenges in building of the FIXME dataset, and we presented our solutions to address them.

\textbf{Challenge.} The consensus in the LAD community is that the availability of RTL data suitable for model training or testing is relatively limited~\cite{liu2024chipnemodomainadaptedllmschip}, and the requirements in the verification domain are more stringent. Using testbench as an example: \textbf{(C1).} There is less data directly usable for verification tasks. We sampled 1,000 Verilog repositories on GitHub, but less than 30\% provided specification, RTL code, and testbench, with varied formats that cannot be easily aggregated for tasks like code generation. \textbf{(C2).} Careful screening is required to ensure initial verification passes: Even for projects meeting the criteria in \textbf{(C1)}, we need to run simulators to double-check whether the code and testbench are properly aligned, a time-consuming process. \textbf{(C3).} Scale Variability: The range of project sizes, from simple scripts to complex systems, requires careful selection to ensure suitability. 
In summary, verification benchmarks belong to the category of few-shot domains with limited data scale, variable quality, and higher requirements, posing unique challenges.

\textbf{Solution.} To address the problem in \textbf{(C1)}, we narrowed our data collection scope to OpenCores~\cite{opencores}, balancing the initial data scale, quality, and type coverage. For the challenges in \textbf{(C2)} and \textbf{(C3)},  we proposed an AI-human collaborative approach tailored for such scenarios. Leveraging the capabilities of large language models, such as tool learning~\cite{zhao2024let} and few-shot learning~\cite{li2024flexkbqa}, we designed a multi-agent data construction system. These models handle the automation of data filtering, initial verification, auto-labeling, and preliminary question generation, while human experts focus on reviewing and enhancing the data quality to accelerate the benchmark development.

\begin{figure*}
\centering
\includegraphics[width=\textwidth]{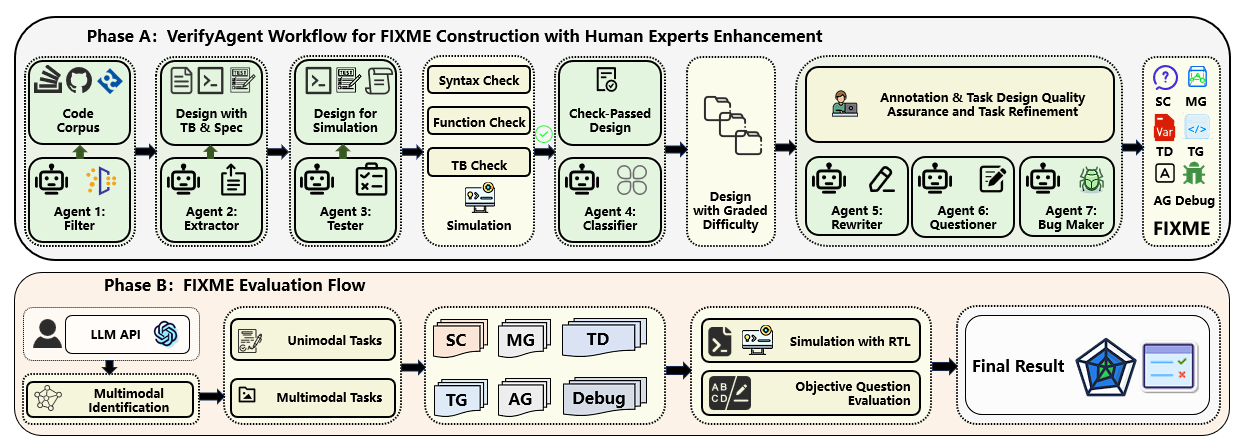}
\caption{Framework of FIXME Construction and Evaluation.}
\label{fig:verifyagent}
\end{figure*}

Specifically for FIXME, we designed the \textbf{VerifyAgent} as shown in Figure~\ref{fig:verifyagent}. Phase A, which we will elaborate on in the section below. It's worth noting that this approach can be applied to other areas of evaluation similar to FV that lack data.

\subsection{Dataset Design}

We started with OpenCores, which includes 485 Verilog projects that have been preliminarily checked by human engineers. These projects cover a wide range of functionalities, from basic arithmetic modules to complex SoCs with tens of thousands of lines of code. More than one-third of the projects have undergone FPGA or silicon verification, providing an acceptable balance between investigation depth and coverage.

To effectively process this data source, we have designed the VerifyAgent framework, which comprises several specialized agents responsible for the various stages of data processing and benchmark construction.

\subsubsection{Automated Basic Data Filtration}
The initial phase involves \textbf{Agent 1}, which systematically scans the project file structures, retaining only those with essential testbench files and verification scripts. Subsequently, \textbf{Agent 2} applies few-shot learning techniques to analyze the selected project directories, extracting the RTL code and testbench files. It then generates Synopsys VCS-compatible verification scripts, which are executed by \textbf{Agent 3} using the VCS task API. Projects that pass these tests become primary candidates for the FIXME benchmark.

\subsubsection{Scale Balancing and Data Review}
To ensure comprehensive coverage and balanced complexity distribution, \textbf{Agent 4} employs a three-tier classification based on Lines of Code (LOC): small-scale (0-100 LOC), medium-scale (100-200 LOC), and large-scale (\textgreater200 LOC) modules.

The classification process distinguishes between leaf modules (standalone modules without instantiations) and hierarchical modules (those containing other module instantiations). For hierarchical modules, we developed a novel decomposition algorithm that expands all contained leaf modules, ensuring accurate complexity grading. This expansion process includes corresponding specification extraction and alignment, maintaining traceability between code and documentation.

Following the automated classification, human experts conduct a double review, applying necessary adjustments based on functional complexity, including control flow complexity, state machine depth, and interface complexity. This hybrid approach ensures that the classification reflects both quantitative and qualitative complexity measures. Subsequently, \textbf{Agent 5} performs semantic-preserving transformation of design specifications through controlled paraphrasing and structural reorganization, effectively mitigating potential model memorization while preserving document accuracy.

\subsubsection{Task Quality Enhancement}
To ensure the accuracy and reliability of the FIXME benchmark, a team of human experts meticulously supplemented the testbench, testcases, and SVAs for each candidate module: 1) optimized testbench achieving \textgreater90\% functional coverage; 2) added exhaustive testcase specifications; 3) designed SVAs for formal verification; and 4) designed the interface for the reference model. This process yielded dual benefits:
\begin{itemize}
\item Identification and elimination of modules exhibiting corner-case failures, resulting in a more robust dataset.
\item Generation of high-fidelity reference solutions across the whole process, minimizing evaluation ambiguity.
\end{itemize}

Quantitatively, our enhancement efforts encompassed: 1) optimization of 58K tokens of technical specifications, 2) rigorous audit of 50K lines of design code and choice of 15K of them, 3) addition of 25K lines of verification testbenches, 4) integration of 1,235 critical assertions, and 5) development of a unified reference model debugging interface.

These improvements resulted in significant coverage improvements, with average line coverage increasing by 7.5\% and toggle coverage improving by 43.14\%, ensuring comprehensive verification of design functionality.

For detailed tasks, there are two kind of tasks that need to be designed based on data. For specification comprehension, \textbf{Agent 6}  generates semantically diverse multiple-choice questions. The questions are structured to evaluate comprehension across different abstraction levels, from basic feature identification to complex behavioral analysis. Concurrently, \textbf{Agent 7} introduces fault injection across various debugging complexity levels, with expert verification ensuring realistic bug scenarios that reflect common hardware design issues.

This AI-human collaborative framework achieved a 95\% reduction in manual processing requirements while maintaining high-quality standards. The resulting FIXME benchmark comprises 30 tasks per verification sub-domain, featuring balanced representation across functionality types and LOC categories, establishing a comprehensive evaluation framework for LLM-aided verification capabilities.

\subsection{Criteria Metric}

To quantitatively assess tasks across verification sub-domains, we define a comprehensive set of evaluation metrics as shown in Table~\ref{Table:example} Column 4.

All of these metrics are expressed as percentages, though their statistical significance varies across different sub-domains.

First, the metric of SC is the pass rate, which evaluates the model’s understanding of fundamental design requirements by measuring the proportion of correctly answered objective questions out of the total number of such questions.

Second, for code generation tasks, syntax and functional pass rate serve as the key evaluation dimensions. Syntax pass rate measures the compliance of the generated code with language syntax rules, while functional pass rate assesses whether the generated output conforms to the expected design specifications. These two metrics are applied in the MG, TG, and AG task categories.

FIXME introduces an assessment based on semantic completeness for the TD sub-domain's short-answer questions. This metric compares the generated content with reference answers by leveraging vector-based semantic similarity. Scores are assigned according to the similarity level: if the similarity is below 60\%, the testcase is considered a failure and assigned a score of zero; if the similarity falls between 60\% and 90\%, the score is proportional to the similarity percentage; if the similarity exceeds 90\%, the testcase is deemed successful.

Specifically for TG tasks, FIXME additionally reports line coverage and toggle coverage. Line coverage quantifies the proportion of executable code lines exercised during verification, while toggle coverage measures the frequency of signal state transitions within the design under test. These metrics collectively reflect the thoroughness and quality of the testbench.

Finally, for Debug tasks, only the simulation pass rate is measured.

Together, these metrics enable FIXME to implement a fine-grained and comprehensive evaluation framework. Except for coverage, all other indicators can be represented by pass rates:

\begin{equation}
\label{eq:Prate}
\text { \textbf{PR} }=\frac{1}{N} \sum_{i=1}^{N} \left[ \Psi'(task_i)=\Psi(task_i) \right] \times 100 \%
\end{equation}

where $\Psi(task_i)$ represents the reference (correct) answer for task $i$, and $\Psi'(task_i)$ denotes the predicted answer. Equation~\ref{eq:Prate} calculates the pass rate of the models across the $N$ tasks in FIXME.

\subsection{Multi-Modal Features}
Unlike most code generation benchmarks, FIXME naturally supports multiple modalities, including text, code, circuit diagrams, tables, and waveform files, to simulate real-world verification scenarios~\cite{chang2024natural,10691753}. For the waveform files required during debugging, we have chosen the VCD format, which can be converted into text-based time-series signal sequences and waveform screenshots for the model to reference during the debugging process.

\subsection{Feedback-Prompted Iteration}
The verification process typically involves iterative refinement based on simulation results. To reflect this, FIXME allows the model to interact with the tools and perform up to three iterations for non-objective questions, providing feedback such as tool run status, error messages, and waveform files to support debugging and adjustment of the model. 

Finally, the evaluation workflow is shown in Figure~\ref{fig:verifyagent}.Phase B. 

\section{EXPERIMENTS}
\label{sec:eval}
\subsection{Experimental Setup}
We used FIXME to rigorously evaluate the capabilities of LLMs in design verification across three difficulty categories and six sub-domains. 

We selected 6 widely-used models: GPT-4o, Claude3, LLaMA3, Gemini-1.5-Pro, Mistral-Large and Semikong~\cite{semikong2024}. This selection represents a diverse set, including SOTA closed-source, open-source, and domain-specific fine-tuned VLSI models. To accommodate the extended context requirements of more complex tasks, we chose versions of these models that support longer input/output token lengths, namely GPT-4o-128k, Claude-3.5-Haiku-200k, LLaMA-3.1-405B-FW-128k, Gemini-1.5-Pro-128k, Mistral-Large-2-128k, and Semikong-70B. The selection of models covers current SOTA closed-source models, open-source models, Mixture-of-Expert models, and VLSI domain-specific models.

To collect comprehensive simulation test coverage data for the testbench generation task, we utilized Synopsys VCS for data collection, focusing on line coverage and toggle coverage metrics. FIXME also supports the collection of additional metrics, such as branch coverage, to provide more detailed data for analysis.

For all models tested, we performed a pass@5 statistical analysis.

\subsection{Evaluation Results}
\begin{figure*}[!t]
    \centering
        \includegraphics[width=\textwidth]{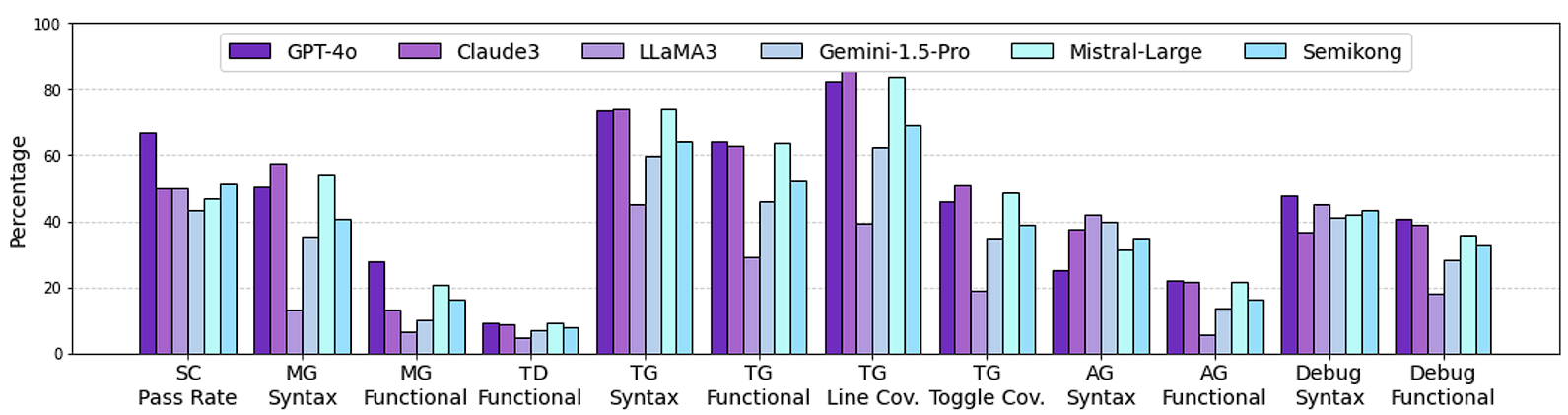}
        \caption{Evaluations of Tested Models on FIXME}
        \label{fig:eval}
\end{figure*}

The final results of our evaluation are presented in Figure~\ref{fig:eval}, and the functional pass rate result is in Table~\ref{lala}. GPT-4o demonstrates the strongest overall verification capabilities, achieving an average functional pass rate of 39.58\%. Meanwhile, Claude3 exhibits a more robust testbench coverage, averaging 68\%. The reference model generation task sees similar performance between the Mistral-Large and GPT-4o. Additionally, LLaMA3 and Semikong maintain comparable syntax correctness to their closed-source counterparts. However, the design of testcases remains a weak area for all the evaluated models. A detailed analysis is provided in the next section.

%TODO：更新Table..

% \usepackage{tabularray}
\begin{table*}
\caption{Results of the Part of Functional Pass Rate. Given the space constraints, we have selected a common evaluation metric across the different sub-domains of the FIXME to present the data and provide the necessary basis for analysis.} %\textcolor{red}{check the  descriptions and examples as I don't have domain knowledge}}
     \vspace{-2pt}
     \label{lala}
\centering
\resizebox{\linewidth}{!}{
\begin{tblr}{
  width = \linewidth,
  colspec = {Q[60]Q[58]Q[37]Q[62]Q[65]Q[62]Q[65]Q[62]Q[65]Q[62]Q[65]Q[62]Q[65]Q[62]Q[65]},
  cell{1}{1} = {r=3}{},
  cell{1}{2} = {r=3}{},
  cell{1}{3} = {r=3}{},
  cell{1}{4} = {c=12}{0.762\linewidth},
  cell{2}{4} = {c=2}{0.127\linewidth},
  cell{2}{6} = {c=2}{0.127\linewidth},
  cell{2}{8} = {c=2}{0.127\linewidth},
  cell{2}{10} = {c=2}{0.127\linewidth},
  cell{2}{12} = {c=2}{0.127\linewidth},
  cell{2}{14} = {c=2}{0.127\linewidth},
  cell{4}{1} = {r=3}{},
  cell{4}{2} = {r=3}{},
  cell{4}{5} = {r=3}{},
  cell{4}{7} = {r=3}{},
  cell{4}{9} = {r=3}{},
  cell{4}{11} = {r=3}{},
  cell{4}{13} = {r=3}{},
  cell{4}{15} = {r=3}{},
  cell{7}{1} = {r=3}{},
  cell{7}{2} = {r=3}{},
  cell{7}{5} = {r=3}{},
  cell{7}{7} = {r=3}{},
  cell{7}{9} = {r=3}{},
  cell{7}{11} = {r=3}{},
  cell{7}{13} = {r=3}{},
  cell{7}{15} = {r=3}{},
  cell{10}{1} = {r=3}{},
  cell{10}{2} = {r=3}{},
  cell{10}{5} = {r=3}{},
  cell{10}{7} = {r=3}{},
  cell{10}{9} = {r=3}{},
  cell{10}{11} = {r=3}{},
  cell{10}{13} = {r=3}{},
  cell{10}{15} = {r=3}{},
  cell{13}{1} = {r=3}{},
  cell{13}{2} = {r=3}{},
  cell{13}{5} = {r=3}{},
  cell{13}{7} = {r=3}{},
  cell{13}{9} = {r=3}{},
  cell{13}{11} = {r=3}{},
  cell{13}{13} = {r=3}{},
  cell{13}{15} = {r=3}{},
  cell{16}{1} = {r=3}{},
  cell{16}{2} = {r=3}{},
  cell{16}{5} = {r=3}{},
  cell{16}{7} = {r=3}{},
  cell{16}{9} = {r=3}{},
  cell{16}{11} = {r=3}{},
  cell{16}{13} = {r=3}{},
  cell{16}{15} = {r=3}{},
  cell{19}{1} = {r=3}{},
  cell{19}{2} = {r=3}{},
  cell{19}{5} = {r=3}{},
  cell{19}{7} = {r=3}{},
  cell{19}{9} = {r=3}{},
  cell{19}{11} = {r=3}{},
  cell{19}{13} = {r=3}{},
  cell{19}{15} = {r=3}{},
  vlines,
  hline{1,4,7,10,13,16,19,22} = {-}{},
  hline{2-3} = {4-15}{},
  hline{5-6,8-9,11-12,14-15,17-18,20-21} = {3-4,6,8,10,12,14}{},
}
Task & Metric    & \centering Level & \centering Model   &         &         &         &         &         &                &         &               &         &          &         \\
           &           &       & \centering GPT-4o  &         & \centering Claude3 &         & \centering LLaMA3  &         & \centering Gemini-1.5-Pro &         & \centering Mistral-Large &         & \centering Semikong &         \\
           &           &       & Result  & Average & Result  & Average & Result  & Average & Result         & Average & Result        & Average & Result   & Average \\
SC         & Pass Rate & L1    & 90.00\% & 67.00\% & 90.00\% & 50.00\% & 70.00\% & 50.00\% & 60.00\%        & 43.33\% & 70.00\%       & 46.67\% & 70.00\%  & 50.00\% \\
           &           & L2    & 60.00\% &         & 40.00\% &         & 50.00\% &         & 40.00\%        &         & 30.00\%       &         & 40.00\%  &         \\
           &           & L3    & 50.00\% &         & 20.00\% &         & 30.00\% &         & 20.00\%        &         & 30.00\%       &         & 40.00\%  &         \\
MG         & Func. PR  & L1    & 32.00\% & 27.94\% & 19.33\% & 13.33\% & 12.00\% & 6.67\%  & 12.00\%        & 10.00\% & 25.00\%       & 20.64\% & 20.00\%  & 16.11\% \\
           &           & L2    & 28.00\% &         & 15.00\% &         & 6.66\%  &         & 8.00\%         &         & 25.33\%       &         & 16.67\%  &         \\
           &           & L3    & 23.82\% &         & 5.66\%  &         & 1.35\%  &         & 10.00\%        &         & 11.57\%       &         & 11.66\%  &         \\
TD        & Func. PR  & L1    & 13.56\% & 11.3\%  & 12.34\% & 9.20\% & 6.54\%  & 4.68\%  & 8.78\%         & 6.79\%  & 11.58\%       & 9.05\%  & 9.68\%   & 7.72\%  \\
           &           & L2    & 11.08\% &         & 10.98\% &         & 5.23\%  &         & 7.45\%         &         & 9.75\%        &         & 8.42\%   &         \\
           &           & L3    & 9.26\%  &         & 4.28\%  &         & 2.27\%  &         & 4.14\%         &         & 5.82\%        &         & 5.07\%   &         \\
TG         & Func. PR  & L1    & 70.15\% & 64.31\% & 71.32\% & 63.02\% & 28.76\% & 29.17\% & 51.00\%        & 46.09\% & 61.23\%       & 63.66\% & 61.23\%  & 52.30\% \\
           &           & L2    & 67.56\% &         & 65.43\% &         & 31.23\% &         & 46.67\%        &         & 71.24\%       &         & 52.35\%  &         \\
           &           & L3    & 55.22\% &         & 52.31\% &         & 27.52\% &         & 40.60\%        &         & 58.51\%       &         & 43.32\%  &         \\
AG        & Func. PR  & L1    & 26.67\% & 21.85\% & 23.33\% & 21.40\% & 7.52\%  & 5.45\%  & 16.67\%        & 13.43\% & 24.53\%       & 22.25\% & 19.78\%  & 16.36\% \\
           &           & L2    & 23.33\% &         & 21.52\% &         & 6.60\%  &         & 13.33\%        &         & 21.54\%       &         & 17.32\%  &         \\
           &           & L3    & 15.55\% &         & 19.35\% &         & 2.23\%  &         & 10.28\%        &         & 20.68\%       &         & 11.99\%  &         \\
Debug      & Func. PR  & L1    & 45.00\% & 40.57\% & 42.56\% & 38.83\% & 24.56\% & 18.05\% & 32.47\%        & 28.44\% & 42.00\%       & 35.70\% & 35.24\%  & 32.61\% \\
           &           & L2    & 39.33\% &         & 37.15\% &         & 19.78\% &         & 27.65\%        &         & 34.53\%       &         & 33.45\%  &         \\
           &           & L3    & 37.37\% &         & 36.78\% &         & 9.82\%  &         & 25.20\%        &         & 30.56\%       &         & 29.15\%  &         
\end{tblr}}
\end{table*}

 \section{DISCUSSION}
 \label{sec:dis}
 \subsection{Result Analysis}
\textbf{1. The effective difficulty gradation is demonstrated by the observed performance trends across various evaluation metrics. }As the LOC and overall task difficulty increase, the pass rates of different models exhibit a declining pattern. This observation aligns with the intended design strategy of FIXME, which aims to sample and construct a benchmark that closely reflects real-world verification scenarios, with appropriate difficulty control to enable meaningful comparisons and the identification of model capability boundaries.

\textbf{2. The multi-faceted evaluation approach has enabled fine-grained insights.} For instance, the GPT-4o model exhibits the strongest performance in specification comprehension tasks, while the Claude3 model leads in syntax pass rate across most tasks and coverage in testbench generation. Furthermore, the Semikong model, which has been fine-tuned for the VLSI domain, demonstrates a 31\% average improvement over its base model, highlighting the impact of targeted fine-tuning strategies.

\textbf{3. The results reveal areas where LLMs encounter significant challenges, such as testcase design and SVA generation. }The low average functional pass rate of 7.72\% across the tested models underscores their limited capability in the intricate reasoning required for formal verification design. Similarly, the need to improve testbench generation coverage, particularly for corner cases and critical signals, is highlighted by the tailored testcase evaluation of the benchmark.

\textbf{4. The observed correlations between different evaluation aspects offer valuable insights for model training strategies.}
It appears that enhancing the verification capabilities of LLMs cannot be accomplished through training exclusively on data from a single stage. Instead, employing synchronous training on datasets from multiple stages of the verification workflow, aligned in an end-to-end manner, may result in more effective improvements.

% Improving the verification capabilities of LLMs may not be achievable through training on data from a single stage alone. Instead, the synchronous training on data from multiple stages of the verification workflow, aligned in an end-to-end fashion, could lead to more effective improvements.

These findings collectively unveil a key pattern: LLMs excel at surface-level tasks involving syntactic generation and basic understanding but struggle with tasks that demand deeper semantic comprehension and abstraction, such as reference model generation, SVA generation, and complex bug debugging. The substantial gaps between syntactic and functional correctness, particularly in model generation and SVA generation, suggest that current LLMs are adept at producing syntactically correct code, but lack the robust understanding required for reliable hardware verification.
Given the existing data scarcity in the FV domain, exploring sustainable data generation, collaborative high-quality dataset construction, and end-to-end design pre-training corpora may represent promising avenues for further research and development.

\subsection{Future Work}
Our work has made significant strides in transitioning from evaluating individual functional verification sub-domains, such as test stimulus generation and SystemVerilog Assertion generation, towards an end-to-end complete functional verification workflow, addressing a critical gap. However, there are still areas for improvement. 

On the one hand, verification represents an important scenario where high-quality data is scarce. While FIXME has proactively open-sourced a total of 100K lines of expert-enhanced evaluation data, there remains a substantial gap that requires more open collaboration. The data construction process employed by FIXME provides a reference approach. On the other hand, from the perspective of modality support, although FIXME is multi-modal, it still needs to consider supporting various types of verification files (e.g., FSDB) encountered in real-world verification workflow and how to process them into an intermediate representation (IR) that is easily understandable by LLMs and encapsulated accordingly.  

Furthermore, considering the emerging applications of LLMs in chip physical design, it may be necessary to evaluate and optimize the potential of LLMs to accelerate design verification across the entire chip creation cycle. This could lead to a paradigm shift in leveraging end-to-end circuit knowledge to enhance verification capabilities of LLMs. 

\section{CONCLUSION}
\label {sec:con}
We presented \textbf{FIXME}, the first comprehensive evaluation framework for assessing the capabilities of language models in the context of hardware functional verification. By carefully constructing 180 real-world-level verification tasks spanning multiple complexity levels, FIXME addresses three critical challenges in LLM-aided verification: the need for a robust suite of evaluation metrics, the requirement for high-quality verification data, and the assessment of complex reasoning capabilities.

We conducted experimental evaluation on SOTA models. While LLMs demonstrate performance in basic tasks, their capabilities decline significantly in scenarios requiring multi-step reasoning or cross-domain integration. The performance degradation follows a consistent pattern, suggesting a fundamental limitation. One of the potential solutions is constructing a hardware domain expertise verification dataset with end-to-end features for fine-grained alignment.

 By identifying specific challenges and providing quantitative metrics for improvement, we hope FIXME can contribute to the development of more reliable AI-powered hardware design methodologies.

\tiny
\balance
\bibliographystyle{unsrt}
\bibliography{reference}

\end{document}